\documentclass[12pt]{article}

\usepackage{amssymb}
\usepackage{latexsym}
\usepackage{epsfig}
\usepackage{citesort}

\def   \btosll   {$b \to s \ell^+ \ell^-$ }
\def   \btoksll  {$B \to K^* \ell^+ \ell^-$ }

\def  \etal      {{\sl et al.} }

\begin{document}


\begin{titlepage}

\renewcommand{\thefootnote}{\fnsymbol{footnote}}

\vspace*{-0.5truecm}
\begin{flushright}
{\tt hep-ph/0505136 \\
YITP-05-21}
\end{flushright}
\vspace*{0.5truecm}
\begin{center}
{\Large \boldmath \bf FB asymmetries in the $B \to K^* \ell^+ \ell^-$
decay: A model independent approach}\\     
\hspace{10pt}\\
\vspace{2.truecm}
{\bf A. S. Cornell$^a$ \footnote{alanc@yukawa.kyoto-u.ac.jp}, 
Naveen Gaur$^b$ \footnote{naveen@physics.du.ac.in} 
and 
Sushil K. Singh$^b$} 
\vskip .8cm
$^a${\sl Yukawa Institute for Theoretical Physics, Kyoto University,
Kyoto 606-8502, Japan.} \\   
\vspace{0.2truecm}
$^b${\sl Department of Physics \& Astrophysics, University of Delhi,
Delhi 110007, India.} \\   
\vspace{2cm}
{\bf Abstract}
\vspace{.2cm}
\begin{quote}
\noindent Among the various ``rare" semi-leptonic decays of $B$-mesons
the \btoksll mode is of special interest. This is because it has the
highest branching ratio among all the semi-leptonic $B$ decays within
the SM. This channel also provides us with a very large number of
possible observables, such as the Forward Backward (FB) asymmetry,
lepton polarization asymmetry etc. Of special interest is the zero
which the FB asymmetry has in this decay mode. In this work we have
studied this zero in the most general model independent framework. 
\end{quote}

\end{center}
\end{titlepage}

\setcounter{footnote}{0}


\section{Introduction \label{intro}}

\par The prospect of not just observing, but quantitatively measuring
physics beyond the Standard Model (SM) at the ``$B$-factories" (BELLE,
BaBar, LHCb etc.) in a matter of years, if not months, has elicited
much excitement in the high-energy phenomenology community. The
primary reason for this is that data from these ``factories" is
already streaming in, whilst other projects capable of probing the
frontiers of known physics may not even commence for several years!
With such a valuable resource at hand it is incumbent upon us to
propose the most experimentally viable tests of any possible
observables of new physics effects. To a large extent this has, since
the experimental observation of the inclusive and exclusive $B \to X_s
\gamma$ and $B \to K^* \gamma$ decays \cite{Alam:1994aw}, been done.  

\par Note that the reason for the flavour-changing neutral current
(FCNC) transitions of $b \to s (d)$ (which occur only through loops in
the SM) having been extensively studied is that these FCNC decays
provide an extremely sensitive test of the gauge structure of the SM
at loop level whilst simultaneously constituting a very suitable tool
for probing new physics beyond the SM. In this quest semi-leptonic
processes have a very crucial role to play, as they are theoretically
and experimentally very clean. Furthermore recall that new physics can
manifest in rare decays through the Wilson coefficients, which can
take values distinctly different from their SM counterparts, or
through possible new structures in the effective Hamiltonian. 

\par However, as the number of possible observables for the initially
observed FCNC processes based on the quark level transition $b \to s
\gamma$ is reasonably small, the study, both experimental and
theoretical, of processes admitting many observables was
required. This led to a focus on observables related to the quark
level process $b \to s(d) \ell^+ \ell^-$, which the phenomenological
community has been studying for quite some time now. Many observables,
such as the FB asymmetry, single and double lepton polarization
asymmetries associated with the final state leptons, have been
studied. Of the decays based on the $b \to s(d) \ell^+ \ell^-$
transition we know that, theoretically, inclusive decays are easier to
calculate, however they are far more difficult to observe than
exclusive decays. On the other hand theoretical predictions of
exclusive decays are model dependent. This model dependence is due to
the fact that in calculating the branching ratios and other
observables for exclusive decays we face the problem of computing the
matrix element of the effective Hamiltonian responsible for the
exclusive decay between the initial and final hadron states. This
problem is related to the non-perturbative sector of QCD and can be
solved only by means of a non-perturbative approach. These matrix
elements have been investigated in the framework of different
approaches, such as chiral theory, relativistic models using the
light-front formalism, effective heavy quark theory and light cone QCD
sum rules
etc. \cite{Aliev:1996hb,Ball:1998kk,Ali:1999mm,Chen:2001ri}. 

\par Many inclusive $B \to X_{s,d} \ell^+ \ell^-$
\cite{Kruger:1996cv,Cornell:2003qt} and exclusive $B \to K (K^*)
\ell^+ \ell^-$ \cite{Aliev:2000jx,Ali:1999mm,Ali:2002jg}, $B \to
\ell^+ \ell^- \gamma$ \cite{RaiChoudhury:2002hf}, $B \to \ell^+
\ell^-$ \cite{Choudhury:1998ze} processes based on $b \to s(d) \ell^+
\ell^-$ have been studied in the literature. But among all these, the
processes $B \to V \ell^+ \ell^-$ are of special interest (where $V$
represents a vector particle). For this reason many of its
observables, such as the FB asymmetry and the lepton polarization
asymmetries, have been extensively studied \cite{Beneke:2001at}. The
FB asymmetry is one of the most important observables in this channel
as various $B$-factories have stated that they shall soon release data
for this asymmetry. The study of the zero of the FB asymmetry in $B
\to V \ell^+ \ell^-$ processes could also be a very useful probe of
the SM. Note that the importance of the vanishing of the FB asymmetry
in $B \to V \ell^+ \ell^-$ was first discussed by Burdman
\cite{Burdman:1998mk} and then in further detail in
\cite{Beneke:2001at} along with elaborations in the technical reports
of various collaborations \cite{Anikeev:2001rk}. In these papers it
was emphasized that within the SM the zero of the FB asymmetry is
largely free from hadronic uncertainties and hence could be a very
useful test. This feature is present in $B \to (K^*, \phi, \rho)
\ell^+ \ell^-$. But amongst all these modes \btoksll has the highest
SM branching ratio. As such this work will analyze the zero of this
asymmetry in the context of \btoksll for a completely model
independent framework assuming the most general form of the effective
Hamiltonian. 

\par It should also be noted that recent results from $B$-factories
are already indicating possible new physics beyond the SM. The
observations regarding $B \to \pi \pi$ and $B \to \pi K$ have already
presented a challenge for theory. Although $B \to \pi \pi$ can be
accommodated within the SM a problem arises as one tries to fix the $B
\to \pi K$ process from the $B \to \pi \pi$ one using SU(3) flavour
symmetry. This problem was first pointed out by Buras \etal
\cite{Buras:2000gc}. Following this a resolution to this {\em puzzle}
was also given by many groups \cite{Buras:2003dj}. It was proposed
that the introduction of a large phenomenological weak phase in the
electroweak penguins could resolve this {\em puzzle}. The presence of
a weak phase in the electroweak penguins has already been studied
extensively in many earlier works
\cite{Buchalla:2000sk,Choudhury:2005rz,Cornell:2004cp,RaiChoudhury:2004pw}.
Results from BELLE and BaBar regarding CP asymmetries in $B \to \eta
K_S$, $B  \to \eta' K_S$, $B \to \phi K_S$ are also hinting at the
presence of some new physics. There have been efforts to resolve these
discrepancies by introducing new set of scalar and pseudo-scalar
operators with complex mass insertions. These have all been
incorporated into our analysis as we try to explore these scenarios,
especially as to how the presence of weak phases in the electroweak
sector and scalar sectors effect the position of the zero of the FB
asymmetry in this decay mode. 

\par As such, the present paper shall be organized along the following
lines: In section 2 we give the most general form of the effective
Hamiltonian and present the analytic results of the branching ratio
and FB asymmetry. We shall then conclude our study in section 3 with a
presentation of our numerical analysis along with our discussion of
these results, summing up with some concluding remarks. 


\section{The Effective Hamiltonian \label{section2}}

\par The processes \btoksll are based on the quark level process
\btosll. The most general model independent (MI) effective Hamiltonian
for this quark level transition can be written as; 
\begin{eqnarray}
{\cal H}_{eff}  &=&  \bigg( \frac{\alpha G}{\sqrt{2}\pi} \bigg)
V_{ts}{V_{tb}}^* \Bigg[   C_{SL}~(\bar{s}i\sigma_{\mu\nu}
\frac{q^{\nu}}{q^2}Lb)~(\bar{\ell}\gamma^{\mu} \ell) + ~
C_{BR}~(\bar{s}i\sigma_{\mu\nu} \frac{q^{\nu}}{q^2}Rb)~(\bar{\ell}
\gamma^{\mu} \ell) \nonumber \\ 
&&+ ~C_{LL}^{tot}
~(\bar{s}_L\gamma_{\mu}b_L)~(\bar{\ell}_L\gamma^{\mu} \ell_L) + ~
C_{LR}^{tot} ~(\bar{s}_L\gamma_{\mu}b_L)~(\bar{\ell}_R \gamma^{\mu}
\ell_R) + ~ C_{RL} (\bar{s}_R\gamma_{\mu}b_R) ~(\bar{\ell}_L
\gamma^{\mu} \ell_L) \nonumber \\ 
&& + ~ C_{RR} (\bar{s}_R\gamma_{\mu}b_R) (\bar{\ell}_R
\gamma^{\mu}\ell_R) + ~ C_{LRLR} (\bar{s}_L b_R) (\bar{\ell}_L \ell_R)
+ ~ C_{LRRL}~(\bar{s}_L b_R)~(\bar{\ell}_R \ell_L) \nonumber \\ 
&& + ~ C_{RLRL} (\bar{s}_R b_L)~(\bar{\ell}_R \ell_L) + ~ C_{RLLR}
(\bar{s}_R b_L)~(\bar{\ell}_L \ell_R) + ~ C_{T}
(\bar{s}\sigma_{\mu\nu}b)~(\bar{\ell} \sigma^{\mu\nu} \ell) \nonumber
\\ 
&& + ~ i C_{TE} \epsilon^{\mu\nu\alpha\beta}
(\bar{s}\sigma_{\mu\nu}b)~(\bar{\ell} \sigma_{\alpha\beta} \ell)
\Bigg] , \label{eq:1} 
\end{eqnarray}
where $L = (1 - \gamma_5)/2$ and $R = (1 + \gamma_5)/2$. $C_X$ denotes
the coefficients of the various four-Fermi interactions, where the
first four of these are already present in the SM. The first two can
be written in terms of the ``standard" SM Wilson coefficients as; 
\begin{eqnarray}
C_{SL} &=& - ~ 2 m_s ~ C^{eff}_7 , \nonumber \\
C_{BR} &=& - ~ 2 m_b ~ C^{eff}_7 .  \nonumber 
\end{eqnarray}
Of the other coefficients two
of these, namely $C_{LL}$ and $C_{LR}$, are also present in the
SM. These can be written in terms of the SM Wilson coefficients as; 
\begin{eqnarray}
C_{LL}^{tot} &=& C^{eff}_9 - C_{10} + C_{LL} , \nonumber \\
C_{LR}^{tot} &=& C^{eff}_9 + C_{10} + C_{LR} .  \nonumber 
\end{eqnarray}
$C_{LL}^{tot}$ and $C_{LR}^{tot}$ are the sum of the contributions
from the SM and any possible new physics. $C_{LL}$ and $C_{LR}$ are
the respective new physics contributions to 
$C_{LL}^{tot}$ and $C_{LR}^{tot}$. 

\par As we are interested in the process \btoksll the observables for
this processes can be calculated from the effective Hamiltonian given
in eqn(\ref{eq:1}). For this we also require the form factors for the
$B \to K^*$ transition. The definition of the $B \to K^*$ form factors
which we will use are given in reference \cite{Ali:1999mm}; 
\begin{eqnarray}
\langle K^*(p_{K^*},\epsilon)|\bar{s}\gamma_\mu (1 \pm
\gamma_5)b|B(p_B) \rangle & = & -\epsilon_{\mu\nu\lambda\sigma}
\epsilon^{*\nu} p^{\lambda}_{K^*} q^{\sigma}
\frac{2V(q^2)}{(m_B+m_{K^*})} \pm~i\epsilon_{\mu}^{*}(m_B+m_{K^*})
\nonumber \\ 
&& \hspace{0.3in} \times A_1(q^2) ~~ \mp~i(p_B+p_{K^*})_{\mu}
(\epsilon^{*} q) \frac{A_2(q^2)}{(m_B+m_{K^*})} \nonumber \\ 
&& \hspace{0.3in} \mp~iq_{\mu} \frac{2m_{K^*}}{q^2} (\epsilon^{*} q)
\bigg[ A_3(q^2)-A_0(q^2) \bigg] , \label{eq:2} \\ 
\langle K^*(p_{K^*},\epsilon)|\bar{s}\sigma_{\mu\nu}b|B(p_B) \rangle &
= & i\epsilon_{\mu\nu\lambda\sigma} \bigg\{ -2
T_1(q^2)\epsilon^{*\lambda} (p_B+p_{K^*})^{\sigma} +\frac{2
(m_B^2-m_{K^*}^2) }{q^2}  \nonumber \\ 
&& \times \bigg(T_1(q^2)-T_2(q^2) \bigg)\epsilon^{*\lambda} q^{\sigma}
-\frac{4}{q^2} \bigg(T_1(q^2)-T_2(q^2)  \nonumber \\ 
&& -\frac{q^2}{(m_B^2-m_{K^*}^2)}~T_3(q^2) \bigg) (\epsilon^{*} q)
p^{\lambda}_{K^*} q^{\sigma} \bigg\} , \label{eq:3} \\ 
\langle K^*(p_{K^*},\epsilon)|\bar{s}i\sigma_{\mu\nu}q^\nu (1 \pm
\gamma_5)b|B(p_B) \rangle & = & 4 \epsilon_{\mu\nu\lambda\sigma}
\epsilon^{*\nu} p^\lambda_{K^*} q^\sigma ~ T_1 (q^2) \nonumber \\ 
&& \pm 2 i \bigg\{ \epsilon_{*\mu}(m_B^2-m_{K^*}^2)-(p_B+p_{K^*})_\mu
(\epsilon^* q) \bigg\}~T_2(q^2) \nonumber \\ 
&& \pm 2 i(\epsilon^* q) \bigg\{ q_\mu -  \frac{(p_B + p_{K^*})_\mu
q^2}{(m_B^2-m_{K^*}^2)} \bigg\} ~ T_3(q^2) , \label{eq:4}  \\ 
\langle K^*(p_{K^*},\epsilon)|\bar{s}(1\pm\gamma_5)b|B(p_B) \rangle &
= & \frac{1}{m_b} \bigg[ \mp2im_{K^*}(\epsilon^{*}q)A_0(q^2) \bigg]
. \label{eq:5} 
\end{eqnarray}

\noindent Using the above definition of the form factors we arrive at
the matrix element for \btoksll as; 
\begin{eqnarray}
{\cal M} & = & \frac{\alpha G}{4\sqrt{2}\pi} V_{ts}{V_{tb}}^* \Bigg[
(\bar{\ell}\gamma^{\mu} \ell) \bigg\{ -2 A
\epsilon_{\mu\nu\lambda\sigma} \epsilon^{*\nu}p_{K^*}^\lambda q^\sigma
- i B \epsilon_\mu^* + i C (\epsilon^* q) ( p_B + p_{K^*})_\mu + i D
(\epsilon^* q) q_\mu \bigg\} \nonumber \\ 
&& \hspace{.3in} +~(\bar{\ell} \gamma^{\mu}\gamma_5 \ell) ~\bigg\{ -2
E \epsilon_{\mu\nu\lambda\sigma} \epsilon^{*\nu}
p_{K^*}^{\lambda}q^{\sigma} - i F\epsilon_{\mu}^{*} + iG(\epsilon^*
q)(p_B+p_{K^*})_{\mu} + i H (\epsilon^* q)q_{\mu} \bigg\} \nonumber \\
&& \hspace{.3in} +~ i Q (\bar{\ell} \ell )~ (\epsilon^* q) + i N
~(\bar{\ell} \gamma_5 \ell)~(\epsilon^* q)   \nonumber \\ 
&& \hspace{.3in} + 16 C_{TE}~(\bar{\ell}\sigma_{\mu\nu} \ell) ~\bigg\{
-2 T_1\epsilon^{*\mu}(p_B+p_{K^*})^{\nu} + B_6\epsilon^{*\mu}q^{\nu} -
B_7(\epsilon^* q)p_{K^*}^{\mu}q^{\nu} \bigg\} \nonumber \\ 
&& \hspace{.3in} +~4 i C_{T}~\epsilon_{\mu\nu\lambda\sigma}
~(\bar{\ell}\sigma^{\mu\nu} \ell) ~\left\{ - 2
T_1\epsilon^{*\lambda}(p_B+p_{K^*})^{\sigma} +
B_6\epsilon^{*\lambda}q^{\sigma} - B_7(\epsilon^*
q)p_{K^*}^{\lambda}q^{\sigma} \right\} \Bigg] , \label{eq:6} 
\end{eqnarray}
where
\begin{eqnarray}
A & = & ( C_{LL}^{tot}+C_{LR}^{tot}+C_{RL}+C_{RR}
)\frac{V}{(m_B+m_{K*})} - 4(C_{BR}+C_{SL})\frac{T_1}{q^2} , 
\nonumber \\ 
B & = & ( C_{LL}^{tot}+C_{LR}^{tot}-C_{RL}-C_{RR} )(m_B+m_{K*})A_1
-4(C_{BR}-C_{SL})(m_{B}^2-m_{K*}^2)\frac{T_2}{q^2} , \nonumber \\ 
C & = & (
C_{LL}^{tot}+C_{LR}^{tot}-C_{RL}-C_{RR})\frac{A_2}{(m_B+m_{K*})}
-4(C_{BR}-C_{SL})
\frac{1}{q^2}\bigg[T_2+\frac{q^2}{(m_{B}^2-m_{K*}^2)}T_3\bigg] ,
\nonumber \\ 
D & = &
2(C_{LL}^{tot}+C_{LR}^{tot}-C_{RL}-C_{RR})m_{K*}\frac{A_3-A_0}{q^2} +
4(C_{BR}-C_{SL})\frac{T_3}{q^2} , \nonumber \\ 
E & = &
(-C_{LL}^{tot}+C_{LR}^{tot}-C_{RL}+C_{RR})\frac{V}{(m_B+m_{K*})} ,
\nonumber \\ 
F & = & (-C_{LL}^{tot}+C_{LR}^{tot}+C_{RL}-C_{RR})(m_B+m_{K*})A_1 ,
\nonumber \\ 
G & = &
(-C_{LL}^{tot}+C_{LR}^{tot}+C_{RL}-C_{RR})\frac{A_2}{(m_B+m_{K*})} ,
\nonumber \\ 
H & = &
2(-C_{LL}^{tot}+C_{LR}^{tot}+C_{RL}-C_{RR})m_{K*}\frac{A_3-A_0}{q^2} ,
\nonumber \\ 
Q & = & -2(C_{LRRL}+C_{LRLR}-C_{RLRL}-C_{RLLR})\frac{m_{K*}}{m_b}A_0 ,
\nonumber \\ 
N & = & -2(C_{LRLR}+C_{RLRL}-C_{RLLR}-C_{LRRL})\frac{m_{K*}}{m_b}A_0 ,
\nonumber \\ 
B_6 & = & 2(m_{B}^2-m_{K*}^2)\frac{T_1-T_2}{q^2} , \nonumber \\
B_7 & = & \frac{4}{q^2}\bigg(T_1-T_2-\frac{q^2}{(m_{B}^2-m_{K*}^2)}T_3
\bigg) . \label{eq:7} 
\end{eqnarray}
The differential decay rate can now be evaluated from the matrix
element given in eqn(\ref{eq:6}) as; 
\begin{equation}
\frac{d \Gamma}{d s} = \frac{G^2 \alpha^2}{2^{14} \pi^5}
|V_{tb}V_{ts}^{*}|^2 v \lambda^{1/2} m_B \Delta , 
\label{eq:8} 
\end{equation}
with
\begin{eqnarray}
\Delta & = & \frac{8}{3} \lambda m_B^4 \hat{s} \bigg\{ (3 - v^2)|A|^2
+ 2 v^2 |E|^2 \bigg\} +\frac{1}{3 \hat{r}} (3 - v^2) \bigg\{ (\lambda
+ 12\hat{r}\hat{s})|B|^2 + \lambda^2 m_B^4 |C|^2 \bigg\} \nonumber \\ 
&&   + \frac{1}{3 \hat{r}} \bigg\{ \lambda (3 - v^2) + 24 \hat{r}
\hat{s} v^2 \bigg\} |F|^2 + \frac{\lambda m_B^4}{3 \hat{r}}
\bigg\{\lambda (3 - v^2) - 3\hat{s}(\hat{s} - 2\hat{r} - 2)(1 - v^2)
\bigg\} |G|^2 \nonumber \\ 
&& + \frac{1}{\hat{r}} \lambda m_B^2 \hat{s} \bigg\{ \hat{s} m_B^2 (1
- v^2)|H|^2 + |N|^2  + v^2|Q|^2 \bigg\} - {2 \over 3} \frac{\lambda
m_B^2}{\hat{r}} \bigg\{ (1 - \hat{r} - \hat{s})(3 - v^2) Re(B C^*)
\nonumber \\ 
&& + \left((1 - \hat{r} - \hat{s})(3 - v^2) + {9 \over 2} \hat{s} (1 -
v^2) \right) Re(F G^*) + 3 \hat{s} (1 - v^2) \bigg( Re(FH^*) - m_B^2
(1 - \hat{r}) \nonumber \\ 
&&  \times Re(GH^*) \bigg) \bigg\} +\frac{4 \lambda m_B}{\hat{r}}
\hat{m}_\ell \bigg\{ m_B^2 \left( (1-\hat{r}) Re(G N^*) + \hat{s} Re(H
N^*) \right) - Re(F N^*) \bigg\} \nonumber \\ 
&&  + 512 \lambda m_B^3 \hat{m}_{\ell} T_1 Re(C_T A^*) + \frac{32
m_B}{\hat{r}} \hat{m}_{\ell} \bigg\{ 2(\lambda+12\hat{r}\hat{s}) B_6 -
\lambda m_B^2 (1-\hat{r}-\hat{s}) B_7 \nonumber \\ 
&& - 4(\lambda + 12\hat{r}(1-\hat{r})) T_1 \bigg\} Re(C_{TE} B^*) +
\frac{32}{\hat{r}} \lambda m_B^3 \hat{m}_\ell \bigg\{
-2(1-\hat{r}-\hat{s}) B_6 + \lambda m_B^2 B_7 \nonumber \\ 
&& + 4(1+3\hat{r}-\hat{s})T_1 \bigg\} Re(C_{TE}C^*)  +
\frac{16}{3\hat{r}}m_{B}^2 \hat{s}\bigg\{ \lambda^2 m_{B}^4 B_7^2 +4(
\lambda +12\hat{r}\hat{s})B_6^2 \nonumber \\ 
&& - 4 \lambda m_B^2 (1 - \hat{r} - \hat{s})B_6 B_7  - 16 \left(
\lambda +12\hat{r}(1- \hat{r}) \right) B_6 T_1 + 8 \lambda m_B^2 (1 +
3\hat{r} - \hat{s}) B_7 T_1 \bigg\} \nonumber \\ 
&& \times \bigg( v^2 |C_T|^2 +4 (3 - 2 v^2) |C_{TE}|^2 \bigg)
+\frac{256}{3 \hat{r}} m_{B}^2 |T_1|^2{|C_T|}^2 \bigg\{ \hat{s} v^2
\bigg(\lambda - 12 \hat{r}(\hat{s}-2 \hat{r}-2 ) \bigg) \nonumber \\  
&&  + 8 \lambda \hat{r}(3 - v^2) \bigg\}  + \frac{1024}{3 \hat{r}}
m_{B}^2 |T_1|^2 |C_{TE}|^2 \bigg\{ 12\lambda \hat{r} + (3 - v^2)
\bigg( \lambda (\hat{s}-8\hat{r})+12\hat{r}(1-\hat{r}^2) \bigg)
\bigg\} , \label{eq:9} 
\end{eqnarray}
\noindent where $\lambda = 1 + \hat{r}^2 + \hat{s}^2 - 2 (\hat{r} +
\hat{s})-2\hat{r} \hat{s}$, $v =
\displaystyle\sqrt{1-\frac{4\hat{m}_\ell^2}{\hat{s}}}$, $ \hat{r} = 
\displaystyle m_{K^*}^2/m_B^2$, $\hat{m}_\ell =
\displaystyle m_\ell/m_B$ and $\hat{s} = \displaystyle 
q^2/m_B^2$. 

The forward backward asymmetry can then be defined as;
\begin{equation}
{\cal A}_{FB} = \frac {\displaystyle \int_{0}^{1} d\cos\theta
\frac{d\Gamma}{ds d\cos\theta} - \int_{-1}^{0} d\cos\theta
\frac{d\Gamma}{ds d\cos\theta} }{\displaystyle \int_{0}^{1}
d\cos\theta \frac{d\Gamma}{ds d\cos\theta} + \int_{-1}^{0} d\cos\theta
\frac{d\Gamma}{ds d\cos\theta} } , \label{eq:10} 
\end{equation}
which in the present case shall give us;
\begin{eqnarray}
{\cal A}_{FB} &=& \frac{v \sqrt{\lambda}}{\Delta} \Bigg[ -8 m_B^2
\hat{s} \bigg\{ Re(B^* E) + Re(A^* F) \bigg\} + \frac{2}{\hat{r}}
m_{B} \hat{m}_\ell \bigg\{ (\hat{r} + \hat{s}-1) {Re}(B^*Q) \nonumber 
\\ 
&& + m_{B}^2 {\lambda} {Re}(C^*Q) \bigg\} - \frac{8}{\hat{r}}
\hat{m}_\ell m_B \bigg\{ \bigg( 2 B_6 (\hat{r} + \hat{s}-1) + B_7
m_{B}^2 {\lambda} + 4 (1 + 11 \hat{r} - \hat{s}) T_1 \bigg) \nonumber
\\ 
&&  \times Re(C_{T}^* F) + m_B^2 \bigg( 2 B_6 (\hat{r} + \hat{s} - 1)
+ B_7 m_{B}^2 \lambda + 4 (1 + 3 \hat{r} - \hat{s}) T_1 \bigg) \bigg(
(\hat{r}-1) Re(C_{T}^*G) \nonumber \\ 
&& - \hat{s} Re(C_{T}^* H) \bigg) \bigg\}  + \frac{4}{\hat{r}} m_B^2
\hat{s} \bigg\{ 2 B_6 (\hat{r} + \hat{s} - 1) + B_7 m_B^2 \lambda + 4
(1 + 3 \hat{r} - \hat{s}) T_1 \bigg\} \nonumber \\ 
&& \bigg\{ 2 Re(C_{TE}^* Q) + Re(C_T^* N) \bigg\} - 256m_B^3
\hat{m}_\ell \bigg\{ B_6 \hat{s} + 2 (\hat{r} - 1) T_1 \bigg\}
Re(C_{TE}^*E) \Bigg] . \label{eq:11} 
\end{eqnarray}


\section{Numerical results and discussion \label{result}}

\par Using the explicit expression of the FB asymmetry given in the
previous section we shall now present our numerical analysis of the
dependence of the zeroes of the FB asymmetry on the various Wilson
coefficients. In our numerical analysis we have used the input
parameters listed in Appendix \ref{appendix:b}. Furthermore, we have
fixed the value of $C_7$ from the results of the $b \to s \gamma$
observation. Note that these results only fix the magnitude of $C_7$
and not the sign, as such we have chosen the SM predicted value of
$C_7 = - 0.313$. Of the remaining SM Wilsons we have chosen the value
of $C_{10} = - 4.56$. And for $C_9^{eff}$ (in the SM), which has both
short distance and long-distance contributions, we have followed the
prescription given in Kr\"{u}ger and Sehgal \cite{Kruger:1996cv}. Note
that the long-distance contributions correspond to the intermediate $c
\bar{c}$ resonances. 

\par In our effective Hamiltonian, given by eqn(\ref{eq:1}), there are
12 coefficients. As pointed out earlier, two of these are related to
$C_7$ by the relation  
$$C_{SL} = - 2 m_b C_7^{eff} ~~~,~~~ C_{BR} = - 2 m_s C_7^{eff} . $$  
Similarly $C_{LL}^{tot}$ and $C_{LR}^{tot}$ can be related to the SM
coefficients $C_9^{eff}$ and $C_{10}$. That is, $C_{LL}^{tot} =
C_9^{eff} - C_{10} + C_{LL}$ and $C_{LR}^{tot} = C_9^{eff} + C_{10} +
C_{LR}$. Note that as the dependence of the zero of the FB asymmetry
on $C_7$ has already been pointed out in earlier works
\cite{Burdman:1998mk}. We have taken the values of $C_{SL}$ and
$C_{BR}$ as fixed by the experimentally measured value of $C_7$ chosen
above. This leaves the ten remaining coefficients as free parameters.

\begin{figure}[ht]
\begin{center}
\epsfig{file=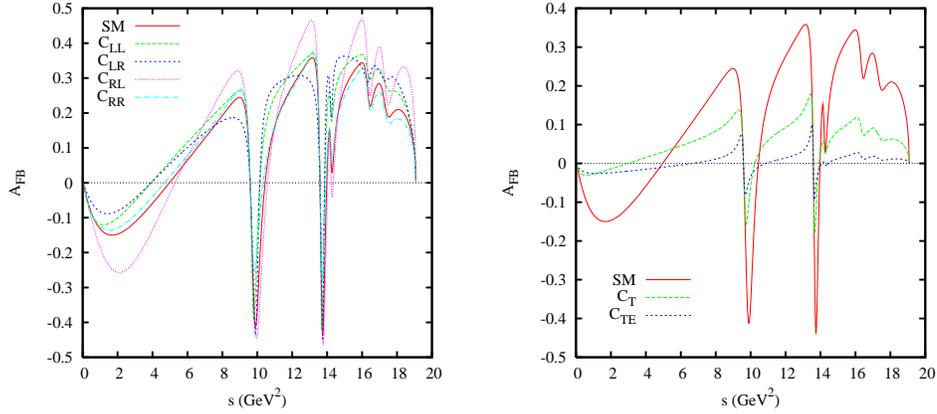,width=.9\textwidth}
\caption{\it The FB asymmetry as a function of the dileptonic
invariant mass ($s$) for various values of the Wilson coefficients. In
the above plots we have chosen $|C_{LL}| = |C_{LR}| = |C_{RL}| =
|C_{RR}| = 3$ and $C_T = C_{TE} = 1.5$.}\label{fig:1} 
\end{center}
\end{figure}
\par As pointed out in section 1, there have been several theoretical 
proposals indicating that some of the new results of the $B$-factories
are pointing towards the presence of weak phases in the electroweak
and scalar sectors. The presence of these phases in the electroweak
sector implies the possibility of $C_{LL}$, $C_{LR}$, $C_{RL}$ and
$C_{RR}$ as being, in general, complex. Similarly the possibility of
the presence of a phase in the scalar sector implies $C_{LRRL}$,
$C_{LRLR}$, $C_{RLLR}$ and $C_{RLRL}$ as also being, in general,
complex. To incorporate these possibilities in our simulations we have
parameterized these coefficients as;  
\begin{eqnarray}
C_X = |C_X| e^{i \phi_X}
\label{eq:12}
\end{eqnarray}
where $X$ can be $LL$, $LR$, $RL$, $RR$, $LRRL$, $LRLR$, $RLRL$ and
$RLLR$.
\begin{figure}[ht]
\begin{center}
\epsfig{file=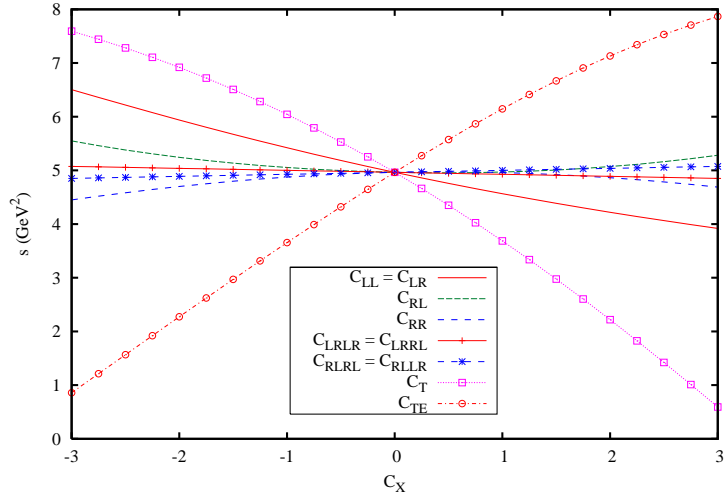,width=.6\textwidth}
\caption{\it The Zeroes of the FB asymmetry as a function of the
Wilsons. The Wilsons here are taken to be real.}\label{fig:2} 
\end{center}
\end{figure}

\par The form factor definitions for $B \to K^*$ which we have used
are given in appendix \ref{appendix:a}. Finally, in our numerical
analysis we have considered only the final state lepton as being the
muon ($\mu$). Our reason for choosing this is due to the extreme
difficulty in detecting an electron in the final state and that the
branching ratio of \btoksll becomes small within the SM for $\tau$ in
the final state. 

\begin{figure}[tb]
\begin{center}
\vskip -.5cm
\epsfig{file=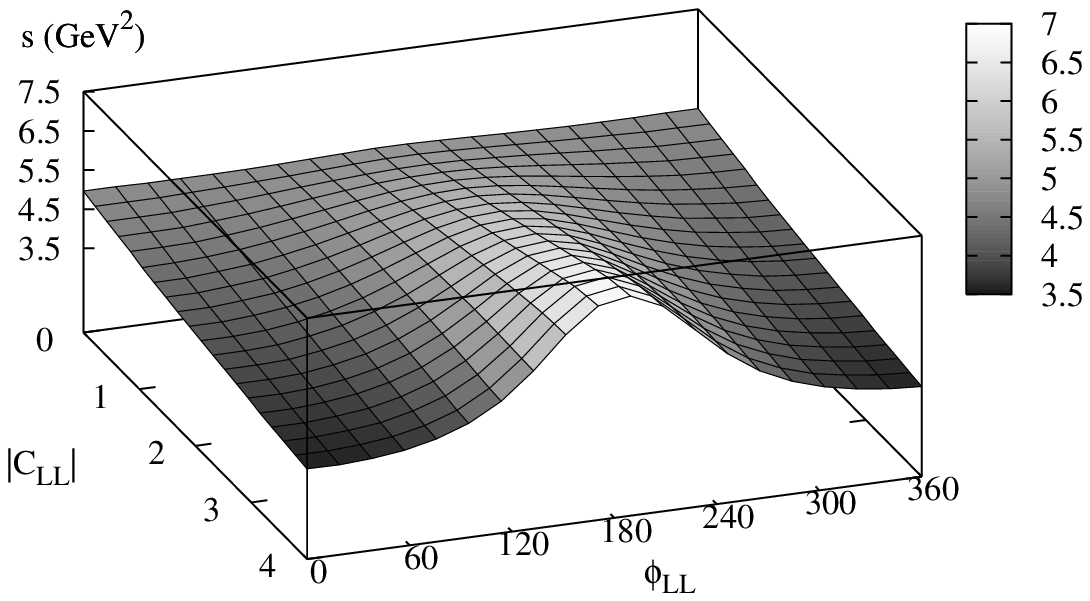,width=.7\textwidth}
\vskip -.5cm
\caption{\it The Zero of the FB asymmetry as a function of the
magnitude and phase of $C_{LL}$.}\label{fig:3} 
\vskip -.5cm
\vskip -.5cm
\epsfig{file=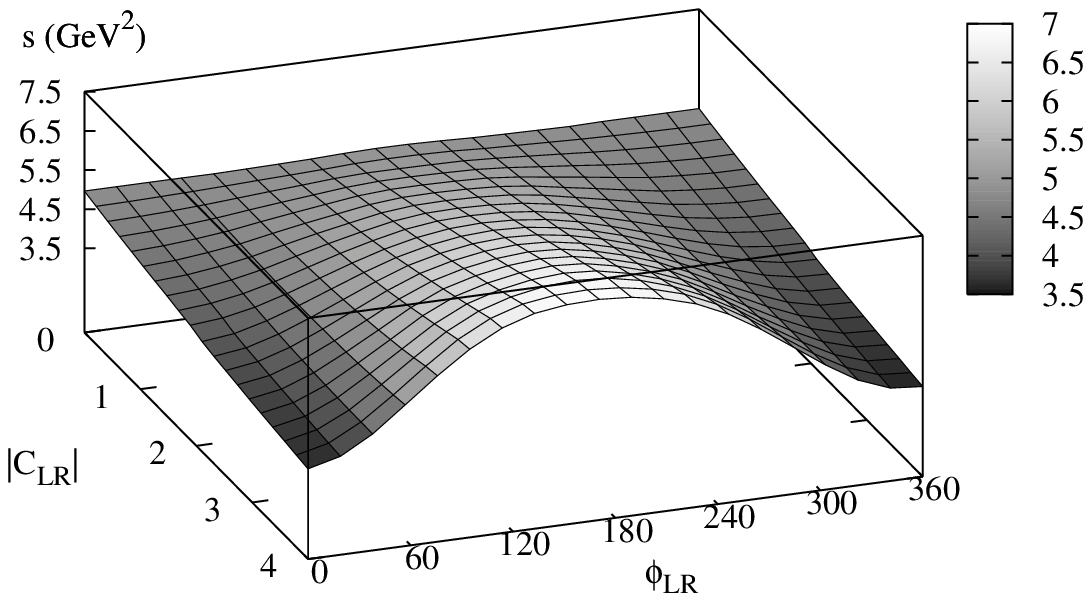,width=.7\textwidth}
\vskip -.5cm
\caption{\it The Zero of the FB asymmetry as a function of the
magnitude and phase of $C_{LR}$.}\label{fig:4} 
\vskip -.5cm
\end{center}
\end{figure}

\par In our first set of graphs, given in Figure \ref{fig:1}, we have
plotted the FB asymmetry as a function of the dilepton invariant mass
for various values of the Wilsons. Our SM value of the zero of the FB
asymmetry is $s = 4.94$. As can be seen from Figure \ref{fig:1} the
value of the zero can be substantially changed for different choices
of the Wilsons. We shall demonstrate this feature further later in
this section. 

\par In Figure \ref{fig:2} we have plotted the zero of the FB
asymmetry as a function of real-valued Wilson coefficients. As can be 
observed from this figure the zero can show substantial modifications,
especially for changes in the tensorial operators, which gives the
greatest change. But as we shall soon see, substantial modifications
to the FB asymmetry zeroes can arise from the other Wilsons when we
include the possible new phases. 

\par The dependence of the zero on both the magnitude and the phase of
the Wilsons was next explored. Note that although all the Wilsons show
changes in the zero we have only shown the results for $C_{LL}$ and
$C_{LR}$, as these two Wilsons show the greatest variation. As such,
in Figure \ref{fig:3} we have shown variation for $C_{LL}$ and in
Figure \ref{fig:4} the dependence of the zero on the magnitude and
phase of $C_{LR}$. From these figures we can not only observe the
dependence of the zero on the magnitude, which further demonstrates
the observations of Figure \ref{fig:2}, but also how it can crucially
depend on the phase of the Wilson. This point can be further clarified
in next set of graphs. 

\begin{figure}[tb]
\vskip -.5cm
\begin{center}
\epsfig{file=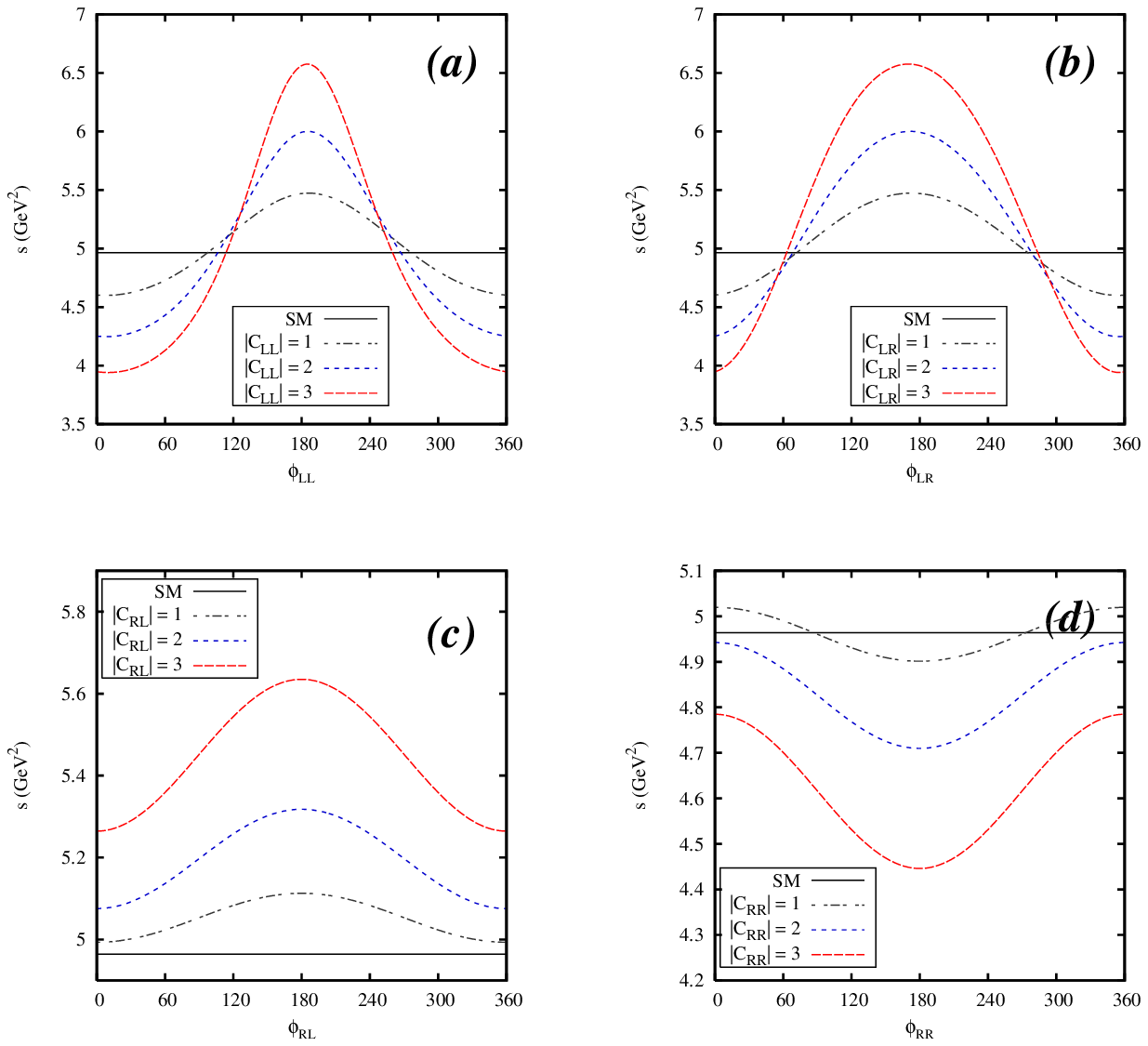,width=.9\textwidth}
\caption{\it Plot of the zero of the FB asymmetry with the phases
of the Wilsons (for different magnitudes of these Wilsons). The 
different panels correspond to {\bf (a)} $C_{LL}$ {\bf (b)} $C_{LR}$ 
{\bf (c)} $C_{RL}$ and {\bf (d)} $C_{RR}$.}\label{fig:5} 
\end{center}
\end{figure}
\begin{figure}[htb]
\begin{center}
\epsfig{file=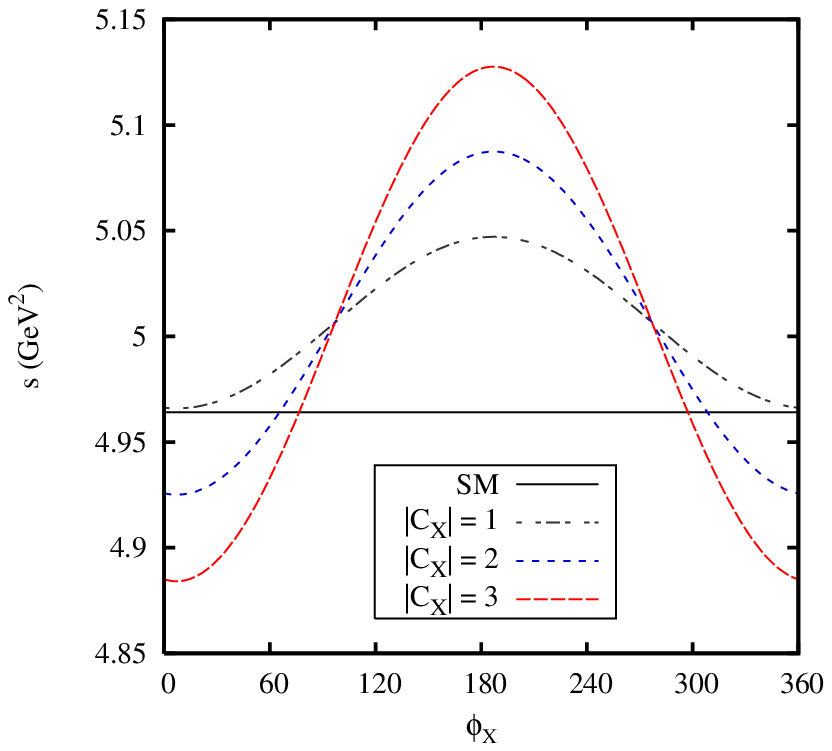,width=.7\textwidth}
\caption{\it Plots of the zero of the FB asymmetry with the phases
of various Wilsons (for different magnitudes of these Wilsons). In the
above figure $X = LRLR, LRRL$.}\label{fig:6} 
\end{center}
\end{figure}
\begin{figure}[htb]
\begin{center}
\epsfig{file=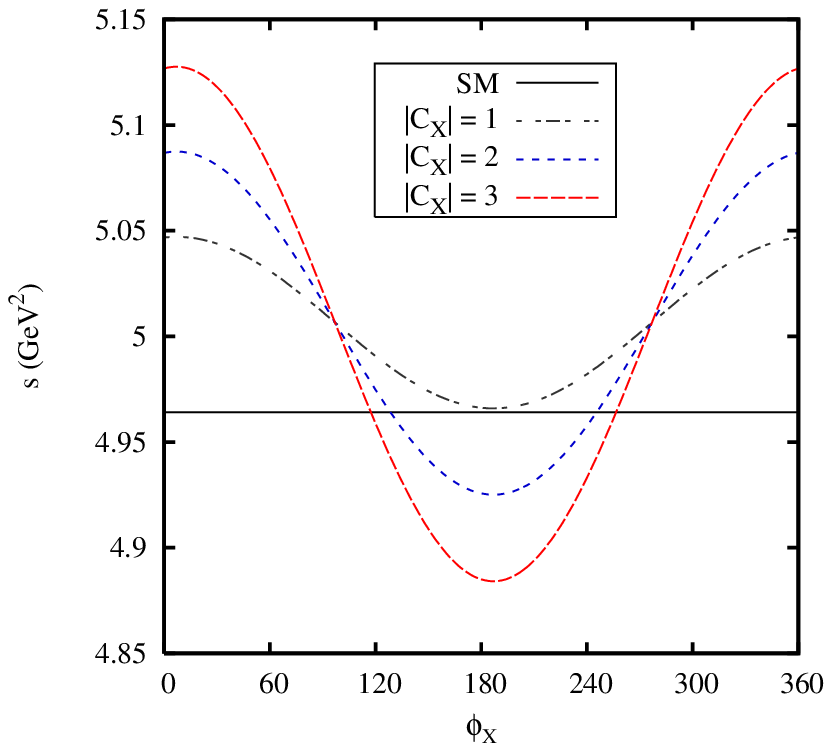,width=.7\textwidth}
\caption{\it Plots of the zero of the FB asymmetry with the phases
of various Wilsons (for different magnitudes of these Wilsons). In the
above figure $X = RLLR, RLRL$.}\label{fig:7} 
\end{center}
\end{figure}

\par In Figure \ref{fig:5} we have plotted the zero of the FB
asymmetry as a function of the phases of the Wilsons for different
magnitudes. This figure emphasizes how strongly the zero of the FB
asymmetry depends on the phase. The variation from the SM result, in
the case of $C_{LL}$ and $C_{LR}$ (which gives us the greatest
variation) can change the asymmetry from 4 to 6.5, a variation of more
than 60\%. In Figure \ref{fig:6} we have plotted the zero as a
function of the phase of $C_{LRLR}$ and $C_{LRRL}$. Similar graphs
have been plotted in Figure \ref{fig:7} for $C_{RLRL}$ and
$C_{RLLR}$. As can be seen from these two figures, coefficients
corresponding to the scalar and electroweak operators do indeed
demonstrate a dependence of the FB asymmetry on the phase. However,
the dependence of the zero in the case of the electroweak operators is
much greater than the scalar operators. 

\begin{figure}[htb]
\begin{center}
\epsfig{file=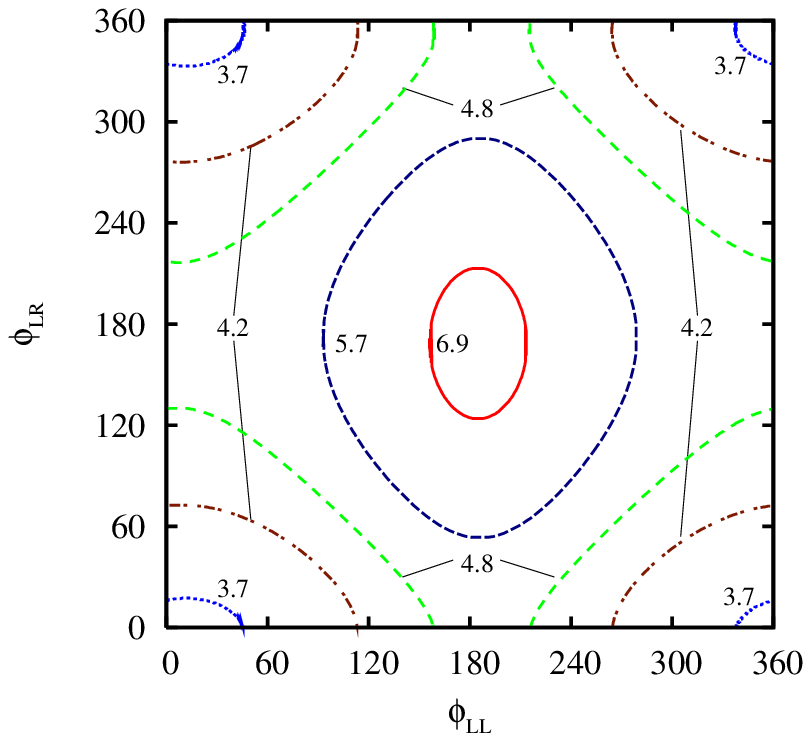,width=.7\textwidth}
\caption{\it Contours plots for the zeroes of the FB asymmetry in the
$\phi_{LL}$ and $\phi_{LR}$ plane. In this plot we have chosen
$|C_{LL}| = |C_{LR}| = 2$.}\label{fig:8} 
\end{center}
\end{figure}
\begin{figure}[htb]
\begin{center}
\epsfig{file=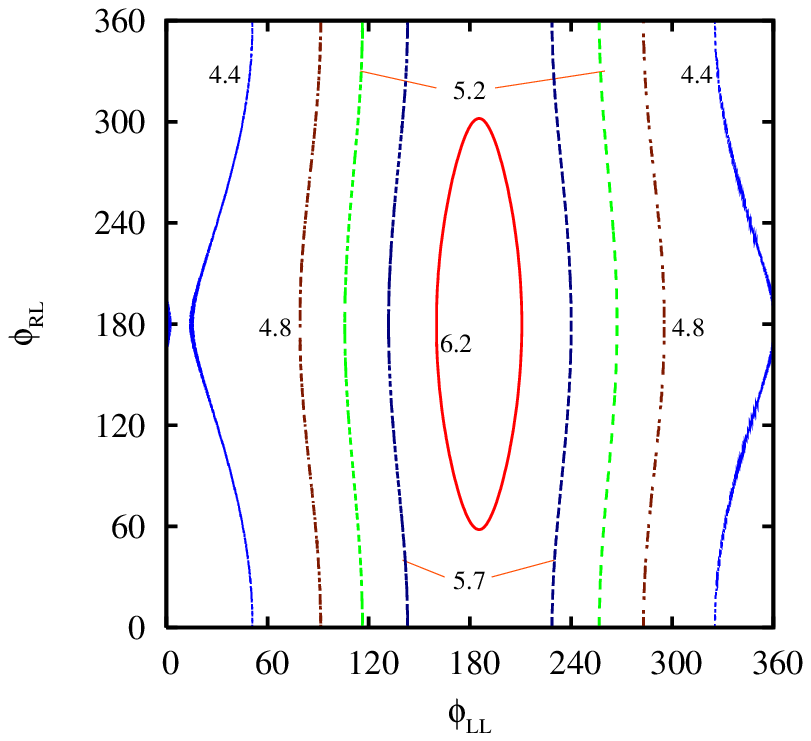,width=.7\textwidth}
\caption{\it Contours plot for the zeroes of the FB asymmetry in the
$\phi_{LL}$ and $\phi_{RL}$ plane. In this plot we have chosen
$|C_{LL}| = |C_{RL}| = 2$.}\label{fig:9} 
\end{center}
\end{figure}
\begin{figure}[htb]
\begin{center}
\epsfig{file=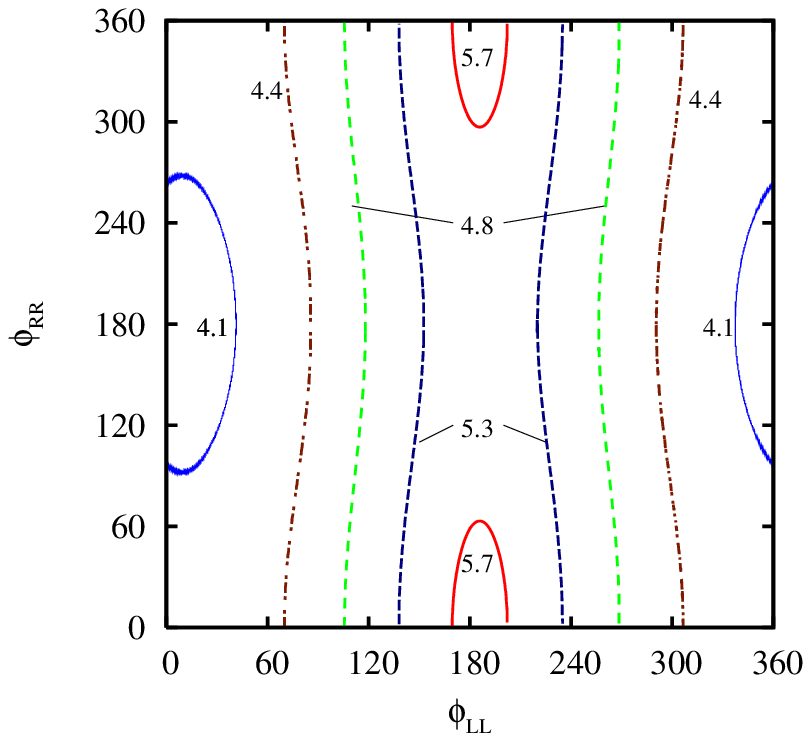,width=.7\textwidth}
\caption{\it Contours plot for the zeroes of the FB asymmetry in the
$\phi_{LL}$ and $\phi_{RR}$ plane. In this plot we have chosen
$|C_{LL}| = |C_{RR}| = 2$.}\label{fig:10} 
\end{center}
\end{figure}

\par To illustrate our previous point further our final set of graphs,
Figures \ref{fig:8}, \ref{fig:9} and \ref{fig:10} show the contour
plots of the zeroes of the FB assuming the presence of two electroweak
operators now having an additional phase. As can be seen from these
graphs the presence of a weak phase in the electroweak sector can give
substantial deviations in the zero of FB asymmetry. 

\par At this point we should like point out that in order to resolve 
the $B \to \pi \pi$ and $B \to \pi K$ {\em puzzle} Buras \etal
\cite{Buras:2000gc} proposed the presence of a phenomenological weak
phase in the electroweak penguins. This phase in effect modifies the
$C_{10}$ Wilson of the SM. This modification not only increases the
magnitude of $C_{10}$, by more than two, but also adds a new large
phase; making the Wilson predominantly imaginary. Note that this kind
of phase will not change the zero of the FB asymmetry, as within the
SM the zero of the FB asymmetry does not depend on $C_{10}$. However,
in general, the presence of extra phases in the electroweak sector
will substantially modify the zero of the FB asymmetry. 

\par From our analysis we have demonstrated that the zero of the FB
asymmetry will not only serve as a valuable test of the SM, as
emphasized in the earlier works \cite{Burdman:1998mk,Beneke:2001at},
but will be a useful probe of any possible new physics.  


\section*{Acknowledgments}
The work of ASC was supported by the Japan Society for the Promotion
of Science (JSPS), under fellowship no P04764. The work of NG was
supported by the Department of Science \& Technology (DST), India,
under grant no SP/S2/K-20/99. 


\appendix


\section{Form Factors \label{appendix:a}}
The form factor definitions which we have used are as given in
Ali. \etal \cite{Ali:1999mm};  
\begin{equation}
F(\hat{s}) = F(0) \mathrm{exp}\left(c_1 \hat{s} + c_2 \hat{s}^2 \right) ,
\label{appendix:a:1}
\end{equation}
where the values of $c_1$ and $c_2$ are given in Table \ref{app:table:1}.

\begin{table}[htb]
\vskip .5cm
\begin{center}
\begin{tabular}{|c | c | c | c | c | c | c | c |} \hline \hline 
 & $A_1$ & $A_2$ & $A_0$ & $V$ & $T_1$ & $T_2$ & $T_3$  \\ \hline 
F(0)  & 0.337 & 0.282 & 0.471 & 0.457 & 0.379 & 0.379 & 0.260 \\
$c_1$ & 0.602 & 1.172 & 1.505 & 1.482 & 1.519 & 0.517 & 1.129 \\
$c_2$ & 0.258 & 0.567 & 0.710 & 1.015 & 1.030 & 0.426 & 1.128 \\
\hline 
\end{tabular}
\caption{\it Form factors for the $B \to K^*$ transition.}\label{app:table:1}
\end{center}
\end{table}


\section{Input parameters \label{appendix:b}}

\begin{center}
$m_t = 176$ GeV ~~,~~ $m_c = 1.4$ GeV ~~,~~ $m_\mu = 0.105$ GeV \\
$m_B = 5.26$ GeV ~~,~~ $m_b = 4.8$ GeV ~~,~~ $sin^2\theta_w = 0.23$
~~,~~ $\alpha = 1/130$ ~~,~~ $m_{K^*} = 0.892$ GeV. 
\end{center}



\end{document}